\documentclass[sn-mathphys-ay]{sn-jnl}

\usepackage{graphicx}%
\usepackage{multirow}%
\usepackage{amsmath,amssymb,amsfonts}%
\usepackage{amsthm}%
\usepackage{mathrsfs}%
\usepackage[title]{appendix}%
\usepackage{xcolor}%
\usepackage{textcomp}%
\usepackage{manyfoot}%
\usepackage{booktabs}%
\usepackage{algorithm}%
\usepackage{algorithmicx}%
\usepackage{algpseudocode}%
\usepackage{listings}%
\usepackage{natbib}

\theoremstyle{thmstyleone}%
\newtheorem{theorem}{Theorem}
\newtheorem{prop}[theorem]{Proposition}%

\theoremstyle{thmstyletwo}%

\theoremstyle{thmstylethree}%

\usepackage{amsmath}
\usepackage{amsfonts}
\usepackage{amssymb}
\usepackage{mathrsfs}
\usepackage{amsthm}
\usepackage{mathtools}

\newcommand{\mbf}[1]{\boldsymbol{#1}}
\newcommand{\mbb}[1]{\mathbb{#1}}
\newcommand{\mcal}[1]{\mathcal{#1}}
\newcommand{\mrm}[1]{\textrm{#1}}

\newcommand{\norm}[1]{\left\|{#1}\right\|}
\newcommand{\abs}[1]{\left|{#1}\right|}

\begin{document}

\title[Genetic Heritability]{Estimating Broad Sense Heritability via Kernel Ridge Regression}


\author[1,2]{\fnm{Olivia} \sur{Bley}}\email{olivia.bley7@gmail.com}
\equalcont{These authors contributed equally to this work.}

\author[1]{\fnm{Elizabeth} \sur{Lei}}\email{elizabethlei72@gmail.com}
\equalcont{These authors contributed equally to this work.}

\author[1,3]{\fnm{Andy} \sur{Zhou}}\email{ andyzhou765@gmail.com}
\equalcont{These authors contributed equally to this work.}

\author*[1]{\fnm{Xiaoxi} \sur{Shen}}\email{rcd67@txstate.edu}

\affil*[1]{\orgdiv{Department of Mathematics}, \orgname{Texas State University}, \orgaddress{\street{601 University Dr.}, \city{San Marcos}, \postcode{78666}, \state{TX}, \country{U.S.A.}}}

\affil[2]{\orgname{San Marcos High School}, \orgaddress{\street{2601 Rattler Road}, \city{San Marcos}, \postcode{78666}, \state{TX}, \country{U.S.A.}}}

\affil[3]{\orgname{William P. Clements High School}, \orgaddress{\street{4200 Elkins Rd.}, \city{Sugar Land}, \postcode{77479}, \state{TX}, \country{U.S.A.}}}


\abstract{The broad-sense genetic heritability, which quantifies the total proportion of phenotypic variation in a population due to genetic factors, is crucial for understanding trait inheritance. While many existing methods focus on estimating narrow-sense heritability, which accounts only for additive genetic variation, this paper introduces a kernel ridge regression approach to estimate broad-sense heritability. We provide both upper and lower bounds for the estimator. The effectiveness of the proposed method was evaluated through extensive simulations of both synthetic data and real data from the 1000 Genomes Project. Additionally, the estimator was applied to data from the Alzheimer’s Disease Neuroimaging Initiative to demonstrate its practical utility.}




\maketitle

\section{Introduction}
\label{s:intro}
Genetic heritability, which measures the proportion of phenotypic variance explained by genetic factors, is a crucial quantity that can enhance our understanding of the genetic mechanisms underlying complex phenotypes \citep{lee2011estimating}. As noted by \citet{zhu2020statistical}, two types of heritability can be estimated. The first is broad-sense heritability ($H^2$), which evaluates the proportion of phenotypic variance explained by all genetic factors, including additive effects, dominance effects, and epistatic interactions. The second type, narrow-sense heritability ($h^2$), represents the proportion of phenotypic variance explained solely by additive genetic effects.

Since the success of Genome-Wide Association Studies (GWAS), hundreds of single nucleotide polymorphisms (SNPs) associated with complex traits have been discovered \citep{hindorff2009potential, donnelly2008progress}. However, SNPs identified through GWAS explain only a small fraction of the genetic variation for any given trait, a well-known issue referred to as "missing heritability" \citep{maher2008personal, manolio2009finding}. One possible explanation is that each individual SNP contributes only a tiny effect to the trait; thus, many SNPs remain undetected as they fail to meet the stringent significance threshold in GWAS \citep{yang2010common}. As a result, existing literature on narrow-sense heritability estimation often relies on modeling the joint effects of multiple SNPs, with linear mixed-effects models (LMMs) playing a key role in constructing narrow-sense heritability estimators. For example, \cite{yang2010common} applied Restricted Maximum Likelihood (REML) multi-variance component analysis on large sample sizes to estimate heritability and uncover genetic architecture. This method demonstrated significant reductions in computation time and memory use compared to previous approaches. The BOLT-REML algorithm revealed that SNP heritability increases with higher-frequency SNPs, while rarer SNPs have a larger effect per allele but explain less variance per SNP. The study further characterized schizophrenia as having a highly polygenic architecture.

In line with this trend, numerous advancements have been made to improve LMM-based narrow-sense heritability estimators. For instance, \cite{lee2011estimating} addressed three main challenges—scale, ascertainment, and SNP quality control—and demonstrated that genetic variation in disease liability, linked to common SNPs in linkage disequilibrium (LD), could be estimated from genome-wide association studies (GWAS) data. First, variance explained on the observed risk scale was transformed to an underlying liability scale. Second, the method accounted for ascertainment, which led to a much higher proportion of cases in the analyzed sample than in the general population. Third, the new approach minimized inflated heritability estimates by preventing artificial differences between case and control genotypes through rigorous GWAS quality control. This method was tested using simulated data and real GWAS data from three Wellcome Trust Case Control Consortium (WTCCC) datasets: Crohn's disease, bipolar disorder, and type I diabetes. Additionally, \citet{speed2012improved} proposed a modified kinship matrix that weights SNPs based on local LD, significantly reducing bias in heritability estimates and enhancing estimation precision.

Several additional approaches have been developed to improve narrow-sense heritability estimation. For example, with the reduced cost of whole-genome sequencing and advances in data storage, biobank data has become accessible. \cite{hou2019accurate} explored an estimator derived under a model that minimizes assumptions about genetic architecture, acknowledging that models with numerous assumptions risk bias due to the unknown nature of genetic architecture. They proposed a generalized random effects (GRE) model that makes minimal assumptions about genetic architecture. They found that the GRE estimator is nearly unbiased across all architectures compared to existing methods, with a maximum bias of around 2\%. As anticipated, the GRE model demonstrated robust performance across architectures, with a runtime of O($N{p_k}^2$) for chromosome $k$ with $p_k$ SNPs, taking approximately 3 hours and 60 GB of memory for 50,000 SNPs. Another direction involves estimating narrow-sense heritability using only summary data, such as the LD matrix. Although individual-level data can provide more precise estimates, its accessibility is often limited by data-sharing restrictions. Recently, \citet{li2023accurate} proposed an iterative procedure called HEELS, which uses summary statistics to solve REML-score equations by adapting Henderson’s algorithm for variance component estimation in LMMs. This approach improves the statistical efficiency of heritability estimators based on summary statistics, achieving performance comparable to that of REML.

Although accurately estimating narrow-sense heritability has received considerable attention in the literature, the true relationship between genetic components and phenotypic traits is unknown, and focusing solely on additive (linear) effects may result in a loss of information. Therefore, it is also important to estimate broad-sense heritability accurately. Mathematically, suppose the relationship between SNPs ($\mbf{Z}$) and a phenotypic trait ($Y$) is given by $Y=g(\mbf{Z})+\epsilon$, where $\epsilon$ is a random error. The broad-sense heritability can then be expressed as
\begin{equation}\label{Eq: Heritability Estimator 1}
H^2=\frac{\mrm{Var}[g(\mbf{Z})]}{\mrm{Var}[g(\mbf{Z})]+\mrm{Var}[\epsilon]}.
\end{equation}
If the function $g$ were known and i.i.d. data pairs $(\mbf{Z}_1, Y_1), (\mbf{Z}_2, Y_2), \ldots, (\mbf{Z}_n, Y_n)$ were observed, a natural estimator for $\mrm{Var}[g(\mbf{Z})]$ would be the sample variance $(n-1)^{-1}\sum_{i=1}^n(g(\mbf{Z}_i)-\bar{g})^2$, with $\bar{g}=n^{-1}\sum_{i=1}^n g(\mbf{Z}_i)$. Similarly, a natural estimator for $\mrm{Var}[\epsilon]$ is its sample counterpart, $n^{-1}\sum_{i=1}^n(Y_i-g(\mbf{Z}_i))^2$. An estimator of $H^2$ can be obtained by replacing the variances with their respective estimators. However, since $g$ is unknown, it must also be estimated. In this paper, this estimation will be achieved through kernel ridge regression.

The remainder of the paper is organized as follows: Section 2 provides the details on our proposed estimator for the broad-sense heritability as well as some theoretical properties of the estimator, followed by some simulation studies whose results are presented in Section 3. A real data analysis of the proposed method for the data from the Alzheimer's Neuroimaging Initiative (ADNI) is given in Section 4 and Section 5 provides a discussion of the proposed method and future work.

\textit{Notations}: Throughout the remainder of the paper, we use bold lower-cased alphabetic letters and bold lower-cased Greek letters to denote non-random vectors. Capitalized alphabetic letters and Greek letters will be used to denote random variables, while bold capitalized alphabetic letters and Greek letters will be used to denote matrices or random vectors. For a matrix $\mbf{A}$, $\norm{\mbf{A}}_{op}$ is used to denote the operator norm, which is the largest singular value of $\mbf{A}$. For a vector $\mbf{a}$, $\norm{\mbf{a}}$ is used to denote the regular Euclidean norm. For a square matrix $\mbf{A}$, $l_1(\mbf{A})\geq l_2(\mbf{A})\geq\cdots\geq l_n(\mbf{A})$ are used to denote the eigenvalues of $\mbf{A}$ and $\norm{\mbf{A}}_F$ and $\norm{\mbf{A}}_{op}$ denote the Frobenius norm and the operator norm of $\mbf{A}$ respectively. 

\section{Methodology}
\label{s:method}
Consider the following semi-parametric model for the population:
\begin{equation}\label{Eq: Population Model}
	Y=g(\mbf{Z})+\epsilon,
\end{equation}
where $\mbf{Z}\in\mbb{R}^p$ is a vector containing $p$ SNPs. Suppose that $g\in\mcal{H}_K$ with $\mcal{H}_K$ being a reproducing kernel Hilbert space associated with the kernel function $K(\cdot,\cdot)$. Now, given the observed data pairs $\{(Y_i,\mbf{Z}_i)\}_{i=1}^n$ from a random sample, the sample version of model (\ref{Eq: Population Model}) is given by
\begin{equation}\label{Eq: Sample Model}
	Y_i=g(\mbf{Z}_i)+\epsilon_i,\quad i=1,\ldots,n.
\end{equation}
Here we assume that $\epsilon_i$ are i.i.d with mean zero and variance $\sigma_\epsilon^2$. Estimation of  $g\in\mcal{H}_K$ can be accomplished via minimizing the penalized least squares loss function
\begin{equation}\label{Eq: PLS}
	\hat{g}=\mrm{argmin}_{g\in\mcal{H}_K}\frac{1}{n}\sum_{i=1}^n(Y_i-g(\mbf{Z}_i))^2+\lambda_n\norm{g}_{\mcal{H}_K}^2.
\end{equation}
Thanks to the Representer Theorem \citep{kimeldorf1970correspondence, kimeldorf1971some}, the general solution for the nonparametric function $g(\cdot)$ in (\ref{Eq: PLS}) can be expressed as
$$
g(\cdot)=\sum_{i=1}^n\alpha_iK(\mbf{Z}_i,\cdot),
$$
so that model (\ref{Eq: Sample Model}), in a matrix-vector form, can be expressed as
\begin{equation}\label{Eq: Sample Model Vector}
	\mbf{Y}=\mbf{K\alpha}+\mbf{\epsilon},
\end{equation}
where $\mbf{K}=[K(\mbf{Z}_i, \mbf{Z}_j)]_{i,j=1}^n$ is a kernel matrix constructed based on the observed data $\mbf{Z}_1,\ldots,\mbf{Z}_n$ and the optimization problem (\ref{Eq: PLS}) becomes
\begin{equation}\label{Eq: PLS vector}
	\hat{\mbf{\alpha}}=\mrm{argmin}_{\mbf{\alpha}}\frac{1}{n}(\mbf{Y}-\mbf{K\alpha})^T(\mbf{Y}-\mbf{K\alpha})+\lambda_n\mbf{\alpha}^T\mbf{K\alpha}.
\end{equation}
Set $Q(\mbf{\alpha})=\frac{1}{n}(\mbf{Y}-\mbf{K\alpha})^T(\mbf{Y}-\mbf{K\alpha})$. Taking derivatives with respect to $\mbf{\alpha}$ gives
\begin{align*}
	\frac{\partial Q}{\partial\mbf{\alpha}} & =\frac{2}{n}\mbf{K}(\mbf{Y}-\mbf{K\alpha})+2\lambda_n\mbf{K\alpha}.
\end{align*}
By letting it equal to $\mbf{0}$ and the solution to the equation $\frac{\partial Q}{\partial\mbf{\alpha}}=\mbf{0}$ is given by
\begin{align*}
	\hat{\mbf{\alpha}} & =\left(\mbf{K}+n\lambda_n\mbf{I}_n\right)^{-1}\mbf{Y}
\end{align*}
We estimate the broad sense heritability by using $\hat{\mbf{\alpha}}$ as given above and using $\hat{\sigma}_\epsilon^2=\frac{1}{n}(\mbf{Y}-\mbf{K}\hat{\mbf{\alpha}})^T(\mbf{Y}-\mbf{K}\hat{\mbf{\alpha}})$. Then the broad sense heritability can be estimated via
\begin{equation}\label{Eq: Herit Estimator}
\hat{H}^2=\frac{\frac{1}{n-1}\hat{\mbf{\alpha}}^T\mbf{K}(\mbf{I}_n-\frac{1}{n}\mbf{J}_n)\mbf{K}\hat{\mbf{\alpha}}}{\frac{1}{n-1}\hat{\mbf{\alpha}}^T\mbf{K}(\mbf{I}_n-\frac{1}{n}\mbf{J}_n)\mbf{K}\hat{\mbf{\alpha}}+\hat{\sigma}_\epsilon^2}.
\end{equation}

Now, we will look at the theoretical properties of the proposed heritability estimator $\hat{H}^2$ by providing asymptotic upper bound and lower bound. For simplicity, we denote $\mbf{g}=[g(\mbf{Z}_1),\ldots,g(\mbf{Z}_n)]^T$, $\bar{\mbf{J}}_n=\frac{1}{n}\mbf{J}_n$ and $\hat{\sigma}_g^2=\frac{1}{n-1}\hat{\mbf{\alpha}}\mbf{K}(\mbf{I}_n-\bar{\mbf{J}}_n)\mbf{K}\hat{\mbf{\alpha}}$. To start with, let us decompose $\hat{\sigma}_g^2$ and $\hat{\sigma}_\epsilon^2$ as follows:
\begin{align*}
    \hat{\sigma}_g^2 & =\frac{1}{n-1}\mbf{Y}^T(\mbf{K}+n\lambda_n\mbf{I}_n)^{-1}\mbf{K}(\mbf{I}_n-\bar{\mbf{J}}_n)\mbf{K}(\mbf{K}+n\lambda_n\mbf{I}_n)^{-1}\mbf{Y}\\
    & =\frac{1}{n-1}(\mbf{g}+\mbf{\epsilon})^T(\mbf{K}+n\lambda_n\mbf{I}_n)^{-1}\mbf{K}(\mbf{I}_n-\bar{\mbf{J}}_n)\mbf{K}(\mbf{K}+n\lambda_n\mbf{I}_n)^{-1}(\mbf{g}+\mbf{\epsilon})\\
    & =\frac{1}{n-1}(I_1^g+I_2^g+I_3^g),
\end{align*}
where
\begin{align*}
    I_1^g & =\mbf{g}^T(\mbf{K}+n\lambda_n\mbf{I}_n)^{-1}\mbf{K}(\mbf{I}_n-\bar{\mbf{J}}_n)\mbf{K}(\mbf{K}+n\lambda_n\mbf{I}_n)^{-1}\mbf{g},\\
    I_2^g & =\mbf{\epsilon}^T(\mbf{K}+n\lambda_n\mbf{I}_n)^{-1}\mbf{K}(\mbf{I}_n-\bar{\mbf{J}}_n)\mbf{K}(\mbf{K}+n\lambda_n\mbf{I}_n)^{-1}\mbf{\epsilon},\\
    I_3^g & =2\mbf{g}^T(\mbf{K}+n\lambda_n\mbf{I}_n)^{-1}\mbf{K}(\mbf{I}_n-\bar{\mbf{J}}_n)\mbf{K}(\mbf{K}+n\lambda_n\mbf{I}_n)^{-1}\mbf{\epsilon}.
\end{align*}
Similarly,
\begin{align*}
    \hat{\sigma}_\epsilon^2 & =\frac{1}{n}(\mbf{Y}-\mbf{K}(\mbf{K}+n\lambda_n\mbf{I}_n)^{-1}\mbf{Y})^T(\mbf{Y}-\mbf{K}(\mbf{K}+n\lambda_n\mbf{I}_n)^{-1}\mbf{Y})\\
    & =\frac{1}{n}\mbf{Y}^T(\mbf{I}_n-\mbf{K}(\mbf{K}+n\lambda_n\mbf{I}_n)^{-1})(\mbf{I}_n-\mbf{K}(\mbf{K}+n\lambda_n\mbf{I}_n)^{-1})\mbf{Y}\\
    & =\frac{1}{n}n^2\lambda_n^2\mbf{Y}^T(\mbf{K}+n\lambda_n\mbf{I}_n)^{-2}\mbf{Y}\\
    & =\frac{1}{n}n^2\lambda_n^2(\mbf{g}+\mbf{\epsilon})^T(\mbf{K}+n\lambda_n\mbf{I}_n)^{-1}(\mbf{g}+\mbf{\epsilon})\\
    & =\frac{1}{n}(I_1^\epsilon+I_2^\epsilon+I_3^\epsilon),
\end{align*}
where
\begin{align*}
    I_1^\epsilon & =n^2\lambda_n^2\mbf{g}^T(\mbf{K}+n\lambda_n\mbf{I}_n)^{-2}\mbf{g}\\
    I_2^\epsilon & =n^2\lambda_n^2\mbf{\epsilon}^T(\mbf{K}+n\lambda_n\mbf{I}_n)^{-2}\mbf{\epsilon}\\
    I_3^\epsilon & =2n^2\lambda_n^2\mbf{g}^T(\mbf{K}+n\lambda_n\mbf{I}_n)^{-2}\mbf{\epsilon}.
\end{align*}

Our first observation is that the cross-product term $\frac{1}{n-1}I_3^g$ and $\frac{1}{n}I_e^\epsilon$ are asymptotically negligible under conditions (C1) and (C2). (C1) assumes that the random errors are sub-Gaussian random variables and (C2) requires that the underlying function $g$ is bounded. These assumptions are common in literature. 

\begin{itemize}
    \item[(C1)] \textbf{(Sub-Gaussian Errors)} $\epsilon_1,\ldots,\epsilon_n$ are i.i.d. sub-gaussian random variables with $\max_i\norm{\epsilon_i}_{\psi_2}\leq C$ for some constant $C>0$.

    \item[(C2)] \textbf{(Bounded Signals)} $\norm{g}_\infty\leq M$ for some $M>0$.
\end{itemize}

\begin{prop}\label{Prop: I3g and I3e}
Under Assumptions (C1) and (C2), given $\mcal{Z}:=\sigma(\mbf{Z}_1,\ldots,\mbf{Z}_n)$,
$$
\frac{1}{n-1}I_3^g=o_p(1), \qquad \frac{1}{n}I_3^\epsilon=o_p(1).
$$
\end{prop}

Since $I_3^g$ and $I_3^\epsilon$ are asymptotically negligible, the asymptotic performance of $\hat{H}^2$ is solely determined by the asymptoci behavior of $I_i^g$ and $I_i^\epsilon$, $i\in\{1,2\}$. We first focus on the asymptotic performance of $I_2^g$ and $I_2^\epsilon$.\\

\begin{prop}\label{Prop: Asymptotics of I2}
    Under Assumptions (C1) and (C2), given $\mcal{Z}:=\sigma(\mbf{Z}_1,\ldots,\mbf{Z}_n)$,
    \begin{align*}
        \frac{1}{n-1}I_2^g & =\frac{\sigma_\epsilon^2}{n-1}\mrm{tr}\left((\mbf{K}+n\lambda_n\mbf{I}_n)^{-1}\mbf{K}(\mbf{I}_n-\bar{\mbf{J}}_n)\mbf{K}(\mbf{K}+n\lambda_n\mbf{I}_n)^{-1}\right)+o_p(1)\\
        \frac{1}{n}I_2^\epsilon & =\frac{\sigma_\epsilon^2}{n}\mrm{tr}\left(n^2\lambda_n^2(\mbf{K}+n\lambda_n\mbf{I}_n)^{-2}\right)+o_p(1).
    \end{align*}
\end{prop}

For $I_1^g$ and $I_1^\epsilon$, since both terms are quadratic form, so they naturally are naturally lower bounded from 0. To provide nontrivial bounds for these quadratic forms, we further impose the following assumptions. Condition (C3) requires that $\mbf{g}$ and $\mbf{v}_1$ are not orthogonal so that $\mbf{g}$ does not lie in the null space of $\mbf{K}$. Similarly, (C3) also requires that neither $\mbf{v}_1$ nor $\mbf{g}$ lies in the null space of $\mbf{J}_n$. Condition (C4) requires that there is sufficient cap between the largest eigenvalue and the second largest eigenvalue of the kernel matrix $\mbf{K}$.

\begin{itemize}
    \item[(C3)] There exist constants $c\in(0,1]$ such that 
    \begin{align*}
    \abs{\mbf{v}_1^T\mbf{1}_n} & \geq c\sqrt{n},\\
    \abs{\mbf{v}_1^T\mbf{g}} & \geq c\norm{\mbf{g}} \mrm{ a.s.},\\
    \abs{\mbf{1}_n^T\mbf{g}} & \geq c\sqrt{n}\norm{\mbf{g}} \mrm{a.s.}
    \end{align*}
    where $\mbf{v}_1$ is the eigenvector corresponding to the largest eigenvalue of $\mbf{K}$.

    \item[(C4)] (\textbf{Spectral Gap}) There exists constant $\alpha\in(\max(2c^2-1,0),1]$ such that
    $$
    \frac{l_1(\mbf{K})}{l_2(\mbf{K})}>\frac{\alpha+1-c^2}{c^2}, \mrm{ a.s.}
    $$
    where $l_2(\mbf{K})$ is the second largest eigenvalue of the kernel matrix $\mbf{K}$.
\end{itemize}

\begin{prop}\label{Prop: Prelim Bounds for 1&g}
    Under Assumptions (C3) and (C4),
    $$
    \abs{\mbf{1}_n^T\mbf{K}(\mbf{K}+n\lambda_n\mbf{I}_n)^{-1}\mbf{g}}\geq\alpha\left(\frac{l_2(\mbf{K})}{l_2(\mbf{K})+n\lambda_n}\right)\sqrt{n}\norm{\mbf{g}},
    $$
    provided that $n\lambda_n\geq\frac{(\alpha+1-c^2)l_1(\mbf{K})l_2(\mbf{K})}{c^2l_1(\mbf{K})-(\alpha+1-c^2)l_2(\mbf{K})}$.\\
\end{prop}

As one can see from Proposition \ref{Prop: Prelim Bounds for 1&g}, to obtain a nontrivial lower bound, besides the conditions (C3) and (C4), one also requires that the regularization parameter $\lambda_n$ should be chosen properly as well. In addition, as a consequence of Proposition \ref{Prop: Prelim Bounds for 1&g}, the quantity $\frac{1}{n}(\mbf{1}_n^T\mbf{K}(\mbf{K}+n\lambda_n)^{-1}\mbf{g})^2$ behaves similarly as $\frac{1}{n}(\mbf{1}_n^T\mbf{g})^2$. This observation is given in the following proposition.\\

\begin{prop}\label{Prop: Bounds for 1&g}
    Under Assumption (C3), (C4) and $n\lambda_n\geq(\alpha+1-c^2)l_1(\mbf{K})l_2(\mbf{K})/[c^2l_1(\mbf{K})-(\alpha+1-c^2)l_2(\mbf{K})]$,
    $$
    \alpha^2\left(\frac{l_1(\mbf{K})}{l_1(\mbf{K})+n\lambda_n}\right)^2\leq \frac{\frac{1}{n}(\mbf{1}_n^T\mbf{K}(\mbf{K}+n\lambda_n\mbf{I}_n)^{-1}\mbf{g})^2}{\frac{1}{n}(\mbf{1}_n^T\mbf{g})^2}\leq c^{-2}\left(\frac{l_1(\mbf{K})}{l_1(\mbf{K})+n\lambda_n}\right)^2\\
    $$
\end{prop}

\noindent
Now we are ready to bound the terms that are not asymptotically negligible to compute $\hat{\sigma}_g^2$ and $\hat{\sigma}_\epsilon^2$. The bounds are provided in the following two propositions.
\begin{prop}\label{Prop: Bounds for I1g and I2g}
    Under Assumption (C1)-(C4) and $n\lambda_n\geq(\alpha+1-c^2)l_1(\mbf{K})l_2(\mbf{K})/[c^2l_1(\mbf{K})-(\alpha+1-c^2)l_2(\mbf{K})]$,
    \begin{align*}
        & \frac{nc^2}{n-1}\left(\frac{l_1(\mbf{K})}{l_1(\mbf{K})+n\lambda_n}\right)^2\left[\frac{1}{n}\norm{\mbf{g}}^2-c^{-4}\left(\frac{1}{n}\mbf{1}_n^T\mbf{g}\right)^2\right]\\
        \leq & \frac{1}{n-1}I_1^g\\
        \leq & \frac{n}{n-1}\left(\frac{l_1(\mbf{K})}{l_1(\mbf{K})+n\lambda_n}\right)^2\left[\frac{1}{n}\norm{\mbf{g}}^2-\alpha^2\left(\frac{l_2(\mbf{K})}{l_2(\mbf{K})+n\lambda_n}\frac{l_1(\mbf{K})+n\lambda_n}{l_1(\mbf{K})}\right)^2\left(\frac{1}{n}\mbf{1}_n^T\mbf{g}\right)^2\right],
    \end{align*}
    and
    \begin{align*}
        \frac{\sigma_\epsilon^2}{n-1}\sum_{i=2}^n\left(\frac{l_i(\mbf{K})}{l_i(\mbf{K})+n\lambda_n}\right)^2+o_p(1)\leq\frac{1}{n-1}I_2^g\leq\frac{(1-c^2)\sigma_\epsilon^2}{n-1}\sum_{i=1}^n\left(\frac{l_i(\mbf{K})}{l_i(\mbf{K})+n\lambda_n}\right)^2+o_p(1).\\
    \end{align*}
\end{prop}

\begin{prop}\label{Prop: Bounds for I1e and I2e}
    Under Assumptions (C1) and (C2), given $\mcal{Z}=\sigma\{\mbf{Z}_1,\ldots,\mbf{Z}_n\}$
    \begin{align*}
        \frac{1}{n}\left(\frac{n\lambda_n}{l_1(\mbf{K})+n\lambda_n}\right)^2\norm{\mbf{g}}^2 & \leq\frac{1}{n}I_1^\epsilon\leq\frac{1}{n}\norm{\mbf{g}}^2\\
        \left(\frac{n\lambda_n}{l_1(\mbf{K})+n\lambda_n}\right)^2\sigma_\epsilon^2+o_p(1) & \leq\frac{1}{n}I_2^\epsilon\leq\sigma_\epsilon^2+o_p(1).\\
    \end{align*}
\end{prop}

\noindent
Putting everything together, we have
\begin{theorem}\label{Thm: Bounds for sigmag and sigmae}
    Under (C1)-(C4) and and $n\lambda_n\geq(\alpha+1-c^2)l_1(\mbf{K})l_2(\mbf{K})/[c^2l_1(\mbf{K})-(\alpha+1-c^2)l_2(\mbf{K})]$,
    \begin{align*}
        \hat{\sigma}_g^2 & \geq c^2\tau_1^2\mrm{Var}[g(\mbf{Z})]+(c^2\tau_1^2-c^{-4})(\mbb{E}[g(\mbf{Z})])^2+\sigma_\epsilon^2\int\left(\frac{x}{x+n\lambda_n}\right)^2\mrm{d}F_{-1}(x)+o_p(1)\\
        \hat{\sigma}_g^2 & \leq \tau_1^2\mrm{Var}[g(\mbf{Z})]+(\tau_1^2-\alpha^2(\tau_2/\tau_1)^2)(\mbb{E}[g(\mbf{Z})])^2+(1-c)^2\sigma_\epsilon^2\int\left(\frac{x}{x+n\lambda_n}\right)^2\mrm{d}F(x)+o_p(1),
    \end{align*}
    and
    \begin{align*}
        \hat{\sigma}_\varepsilon^2 & \geq (1-\tau_1)^2\left(\mrm{Var}[g(\mbf{Z})]+\sigma_\varepsilon^2\right)+(1-\tau_1)^2(\mbb{E}[g(\mbf{Z})])^2+o_p(1)\\
        \hat{\sigma}_\varepsilon^2 & \leq\mrm{Var}[g(\mbf{Z})]+\sigma_\varepsilon^2+(\mbb{E}[g(\mbf{Z})])^2+o_p(1),
    \end{align*}
    where
    $$
    \tau_i=\lim_{n\to\infty}\frac{l_i(\mbf{K})}{l_i(\mbf{K})+n\lambda_n}, \quad i=1,2
    $$
    $F(x)$ is the limiting empirical spectral distribution (ESD) of $\mbf{K}$, and $F_{-1}(x)$ is the limiting empirical spectral distribution (ESD) of $\mbf{K}$ by removing the largest eigenvalue of $\mbf{K}$.\\
\end{theorem}

As a result, the bounds for $\hat{\sigma}_\epsilon^2$ and $\hat{\sigma}_g^2$ further provide an upper bound and a lower bound for $\hat{\sigma}_\epsilon^2/\hat{\sigma}_g^2$, which is an important quantity for $\hat{H}^2$ since $\hat{H}^2=1/(1+\hat{\sigma}_\epsilon^2/\hat{\sigma}_g^2)$. In addition, by careful choice of the regularization parameter $\lambda_n$, it can be shown that the bounds provided covers the true value $\sigma_\epsilon^2/\sigma_g^2$.

\begin{prop}\label{Prop: Bounds for sigmae^2/sigmag^2}
    Under the same conditions as in Theorem \ref{Thm: Bounds for sigmag and sigmae},
    \begin{align*}
        & \frac{(1-\tau_1)^2(\mrm{Var}[g(\mbf{Z})]+\sigma_\epsilon^2)+(1-\tau_1)^2(\mbb{E}[g(\mbf{Z})])^2}{c^2\tau_1^2\mrm{Var}[g(\mbf{Z})]+(\tau_1^2-\alpha^2(\tau_2/\tau_1)^2)(\mbb{E}[g(\mbf{Z})])^2+(1-c)^2\sigma_\epsilon^2\int\left(\frac{x}{x+n\lambda_n}\right)^2\mrm{d}F(x)}\\
        \leq & \lim_{n\to\infty}\hat{\sigma}_\epsilon^2/\hat{\sigma}_g^2\\
        \leq & \frac{\mrm{Var}[g(\mbf{Z})]+\sigma_\epsilon^2+(\mbb{E}[g(\mbf{Z})])^2}{c^2\tau_1^2\mrm{Var}[g(\mbf{Z})]+(c^2\tau_1^2-c^{-4})(\mbb{E}[g(\mbf{Z})])^2+\sigma_\epsilon^2\int\left(\frac{x}{x+n\lambda_n}\right)^2\mrm{d}F_{-1}(x)}
    \end{align*}
    Consequently, if the regularization parameter $\lambda_n$ is chosen such that
    \begin{align*}
        \frac{(\alpha+1-c^2)l_1(\mbf{K})l_2(\mbf{K})}{c^2l_1(\mbf{K})-(\alpha+1-c^2)l_2(\mbf{K})} & \leq n\lambda_n\\
        \int\left(\frac{x}{x+n\lambda_n}\right)^2\mrm{d}F_{-1}(x) & \leq\frac{\sigma_g^2}{\sigma_\epsilon^2}(1-c^2\tau_1^2)\\
        \int\left(\frac{x}{x+n\lambda_n}\right)^2\mrm{d}F(x) & \geq\frac{1}{1-c^2}\left[\frac{\sigma_g^4}{\sigma_\epsilon^4}+(1-c^2\tau_1^2)\frac{\sigma_g^2}{\sigma_\epsilon^2}\right.\\
        & \qquad\qquad\left.+\left(\frac{\sigma_g^2}{\sigma_\epsilon^4}-\frac{\tau_1^2-\alpha^2(\tau_2/\tau_1)^2}{\sigma_\epsilon^2}\right)(\mbb{E}[g(\mbf{Z})])^2\right],
    \end{align*}
    then the asymptotic bounds for $\hat{\sigma}_\epsilon^2/\hat{\sigma}_g^2$ cover the true value $\sigma_\epsilon^2/\sigma_g^2$ with $\sigma_g^2=\mrm{Var}[g(\mbf{Z})]$.
\end{prop}

\section{Simulation}
To evaluate the performance of the proposed estimators, we started by creating populations, from which response variables were generated based on the following model:
\begin{equation}\label{Eq: Simulation Model}
Y_i=g(\mbf{Z}_i)+\epsilon_i,\quad i=1,\ldots,N,
\end{equation}
where $N$ is the population size specified in the following subsections; $\epsilon_i$ are i.i.d. normal random variables with mean 0 and standard deviation 0.5. 
The genetic variants $\mbf{Z}_i\in\mbb{R}^p$ were generated under two different scenarios: (1). The genetic variants simulated based on Hardy-Weinberg equilibrium (HWE) and (2). The genetic variants sampled from the 1000-Genome Project \citep{altshuler2010map}. Three underlying functions were used in the simulation studies:
\begin{enumerate}
    \item Linear function: $g(\mbf{Z}_i)=2\mbf{Z}_i^T\mbf{\beta}+5$.
    \item Quadratic function: $g(\mbf{Z}_i)=(\mbf{Z}_i^T\mbf{\beta})^2$.
    \item Trigonometric function: $g(x\mbf{Z}_i)=\sin(\mbf{Z}_i^T\mbf{\beta})+2\mbf{Z}_i^T\mbf{\beta}$.
\end{enumerate} 
The genetic effect $\mbf{\beta}\sim\mcal{N}_p(\mbf{0},\sigma_g^2\mbf{I}_n)$ and the variance component $\sigma_g$ is chosen so that the population broad-sense heritability lies between 0.2 and 0.8. In each scenario, we considered two cases: the low-dimensional case ($N>p$) and the high-dimensional case ($N<p$). We now explain in detail how the $\mbf{Z}_i$'s were simulated in these two scenarios.

\subsection{Scenario 1: Genetic Variants Simulated from HWE}
In this scenario, we started by simulating the minor allele frequencies (MAFs) of the SNPs from the uniform distribution $\mcal{U}[0.01, 0.5]$. Then, Hardy-Weinberg equilibrium (HWE) was applied to simulate the additive genotypes. Specifically, the $j$th SNP with MAF, $MAF_j$, has values 0, 1, and 2 with the following probabilities respectively:
\begin{align*}
    \mbb{P}(SNP_j=0) & =(1-MAF_j)^2,\\
    \mbb{P}(SNP_j=1) & =2MAF_j(1-MAF_j),\\
    \mbb{P}(SNP_j=2) & =MAF_j^2.
\end{align*}
The choices of $N$, $p$, and $\sigma_g$ under different simulation settings are summarized in Table \ref{tab: choice of N and p based on HWE}.
\begin{table}[htbp]
    \centering
    \caption{Choices of $N$, $p$ and $\sigma_g$ for the scenario where the genetic variants were simulated based on HWE.}\label{tab: choice of N and p based on HWE}
    \begin{tabular}{c|ccc|ccc}
    \hline
    & \multicolumn{3}{c|}{Low-dimensional ($N>p$)}
    & \multicolumn{3}{c}{High-dimensional ($N<p$)} \\
    \hline
    & $N$ & $p$ & $\sigma_g$ & $N$ & $p$ & $\sigma_g$\\
    \hline
    Linear & 1000 & 500 & 0.02 & 500 & 1000 & 0.01 \\
    Quadratic & 1000 & 500 & 0.03 & 500 & 1000 & 0.02\\
    Trigonometric & 1000 & 500 & 0.02 & 500 & 1000 & 0.05\\
    \hline
    \end{tabular}
    
\end{table}

\subsection{Scenario 2: Genetic Variants Simulated from Real Genetic Data}
To mimic the real structure of sequence data, genetic variants in this scenario were sampled from the real sequencing data of Chromosome 17: 7,344,328–8,344,327 from the 1000 Genomes Project. The distribution of minor allele frequencies (MAFs) of single nucleotide polymorphisms (SNPs) in this region is heavily skewed to the right and ranges from 0.046\% to 49.954\%. In this scenario, $p$ SNPs were randomly sampled from all SNPs in the sequencing data.  The choices of $N$, $p$ and $\sigma_g$ are summarized in Table \ref{tab: choice of N and p based on 1000 genome}
\begin{table}[htbp]
    \centering
    \caption{Choices of $N$, $p$ and $\sigma_g$ for the scenario where the genetic variants were simulated based on real genetic data.}\label{tab: choice of N and p based on 1000 genome}
    \begin{tabular}{c|ccc|ccc}
    \hline
    & \multicolumn{3}{c|}{Low-dimensional ($N>p$)}
    & \multicolumn{3}{c}{High-dimensional ($N<p$)} \\
    \hline
    & $N$ & $p$ & $\sigma_g$ & $N$ & $p$ & $\sigma_g$\\
    \hline
    Linear & 1092 & 500 & 0.02 & 1092 & 1500 & 0.01 \\
    Quadratic & 1092 & 500 & 0.03 & 1092 & 1500 & 0.015\\
    Trigonometric & 1092 & 500 & 0.05 & 1092 & 1500 & 0.05\\
    \hline
    \end{tabular}
    
\end{table}

\noindent
In both scenarios, the linear kernel matrix $\mbf{K}=p^{-1}\mbf{ZZ}^T$, the polynomial kernel matrix $\mbf{K}=(\mbf{J}_n+p^{-1}\mbf{ZZ}^T)^{\circ 2}$, where $\circ 2$ means elementwise squaring, and the Gaussian kernel $\mbf{K}_{ij}=\exp\left\{-\frac{1}{2}\norm{\mbf{Z}_i-\mbf{Z}_j}^2\right\}$ were used to estimate the heritability. In terms of the regularization parameter $\lambda_n$ in the kernel ridge regression, we predefined a candidacy set $\mcal{C}=\{0.1, 0.5, 0.8, 1, 1.3, 1.5, 2, 2.3,	2.5, 3,	5\}$ for the values of $n\lambda_n$ and calculated the heritability estimator using each candidate. In the low-dimensional case, samples of sizes 600, 700, 800, 900, and 1000 were randomly selected from the corresponding population, while in the high-dimensional case, samples of sizes 100, 200, 300, 400, and 500 were randomly picked from the corresponding population. We then calculated the heritability estimator for each sample size. Such a process was repeated 500 times. 

Based on the results, the polynomial kernel provides the most accurate estimation of the underlying heritability across all settings when the value of $n\lambda_n$ is appropriately chosen. Table \ref{tab: simulation result linear ld} and Table \ref{tab: simulation result linear hd} summarize the mean heritability estimates and standard deviations across 500 repetitions when the underlying function is linear. The results for the other two functions used in the simulations are presented in Appendix B. Here, we only present tables for the three values of $n\lambda_n$ that yield the most accurate estimates for the polynomial kernel.

\begin{table}[htbp]
    \centering
    \caption{Mean and standard deviation of the heritability estimators obtained after running 500 repetitions when the underlying function is linear and the low dimensional case was applied. The true heritability for the HWE scenario is 0.769 and the true heritability for the 1000 Genome Project case is 0.774.}\label{tab: simulation result linear ld}
    \begin{tabular}{c|ccc|ccc}
         \hline
 & \multicolumn{3}{c|}{ Hardy Weinberg Equilibrium}
 & \multicolumn{3}{c}{ 1000 Genome Project} \\
\hline
 & \multicolumn{3}{c|}{$n\lambda_n$}
 & \multicolumn{3}{c}{$n\lambda_n$} \\
\hline
 Linear & 2.3 & 2.5 & 3.0 & 1.5 & 2.0 & 2.3 \\
\hline
 600 & 0.034 (0.002) & 0.031 (0.002) & 0.025 (0.002) & 0.054 (0.005) & 0.042 (0.004) & 0.037 (0.004) \\
 700 & 0.030 (0.002) & 0.027 (0.002) & 0.022 (0.001) & 0.047 (0.004) & 0.037 (0.003) & 0.033 (0.003) \\
 800 & 0.024 (0.001) & 0.022 (0.001) & 0.018 (0.001) & 0.040 (0.003) & 0.031 (0.003) & 0.028 (0.002) \\
 900 & 0.017 (0.001) & 0.016 (0.001) & 0.013 (0.001) & 0.031 (0.002) & 0.025 (0.002) & 0.022 (0.002) \\
 1000 &	0.009 (0.000) & 0.008 (0.000) & 0.007 (0.000) & 0.024 (0.002) & 0.019 (0.001) & 0.017 (0.001) \\
\hline
 Polynomial & 2.3 & 2.5 & 3 & 1.5 & 2.0 & 2.3 \\
\hline
 600 & 0.702 (0.011) & 0.674 (0.012) & 0.608 (0.013) & 0.803 (0.014) & 0.765 (0.016) & 0.745 (0.017) \\
 700 & 0.712 (0.009) & 0.685 (0.009) & 0.622 (0.011) & 0.806 (0.012) & 0.770 (0.013) & 0.750 (0.014) \\
 800 & 0.721 (0.007) & 0.695 (0.008) & 0.635 (0.008) & 0.809 (0.010) & 0.773 (0.011) & 0.755 (0.010) \\
 900 & 0.729 (0.005) & 0.704 (0.006) & 0.646 (0.006) & 0.810 (0.007) & 0.777 (0.008) & 0.758 (0.009) \\
 1000 &	0.736 (0.000) & 0.713 (0.000) & 0.656 (0.000) & 0.811 (0.005) & 0.780 (0.005) & 0.762 (0.006) \\
\hline
 Gaussian & 2.3 & 2.5 & 3 & 1.5 & 2.0 & 2.3 \\
\hline
 600 & 0.265 (0.009) & 0.257 (0.009) & 0.242 (0.009) & 0.423 (0.016) & 0.412 (0.015) & 0.408 (0.015) \\
 700 & 0.271 (0.008) & 0.262 (0.008) & 0.246 (0.007) & 0.426 (0.013) & 0.413 (0.012) & 0.409 (0.012) \\
 800 & 0.275 (0.006) & 0.266 (0.006) & 0.250 (0.005) & 0.428 (0.011) & 0.414 (0.010) & 0.409 (0.009) \\
 900 & 0.281 (0.004) & 0.270 (0.004) & 0.253 (0.004) & 0.430 (0.008) & 0.416 (0.008) & 0.411 (0.007) \\
 1000 &	0.285 (0.000) & 0.275 (0.000) & 0.256 (0.000) & 0.432 (0.005) & 0.417 (0.005) & 0.411 (0.005) \\
\hline
    \end{tabular}
\end{table}

\begin{table}[htbp]
    \centering
    \caption{Mean and standard deviation of the heritability estimators obtained after running 500 repetitions when the underlying function is linear and the high dimensional case was applied. The true heritability for the HWE scenario is 0.594 and the true heritability for the 1000 Genome Project case is 0.719.}\label{tab: simulation result linear hd}
    \begin{tabular}{c|ccc|ccc}
\hline
 & \multicolumn{3}{c|}{ Hardy Weinberg Equilibrium}
 & \multicolumn{3}{c}{ 1000 Genome Project} \\
\hline
 & \multicolumn{3}{c|}{$n\lambda_n$}
 & \multicolumn{3}{c}{$n\lambda_n$} \\
\hline
 Linear & 2.0 & 2.3 & 2.5 & 1.3 & 1.5 & 2.0 \\
\hline
 100 & 0.014 (0.002) & 0.012 (0.002) & 0.010 (0.002) & 0.056 (0.010) & 0.041 (0.008) & 0.034 (0.007) \\
 200 & 0.018 (0.002) & 0.014 (0.002) & 0.013 (0.001) & 0.055 (0.007) & 0.039 (0.005) & 0.034 (0.005) \\
 300 & 0.017 (0.001) & 0.014 (0.001) & 0.012 (0.001) & 0.052 (0.006) & 0.038 (0.005) & 0.033 (0.005) \\
 400 & 0.012 (0.001) & 0.010 (0.001) & 0.009 (0.001) & 0.049 (0.005) & 0.036 (0.004) & 0.031 (0.004) \\
 500 & 0.004 (0.000) & 0.004 (0.000) & 0.003 (0.000) & 0.044 (0.005) & 0.033 (0.003) & 0.028 (0.003) \\
\hline
 Polynomial & 2.0 & 2.3 & 2.5 & 1.3 & 1.5 & 2.0 \\
\hline
 100 & 0.682 (0.013) & 0.618 (0.015) & 0.576 (0.015) & 0.766 (0.046) & 0.688 (0.056) & 0.643 (0.059) \\
 200 & 0.700 (0.011) & 0.640 (0.013) & 0.601 (0.013) & 0.779 (0.031) & 0.708 (0.037) & 0.670 (0.039) \\
 300 & 0.708 (0.009) & 0.650 (0.010) & 0.614 (0.011) & 0.783 (0.024) & 0.717 (0.028) & 0.683 (0.031) \\
 400 & 0.714 (0.006) & 0.658 (0.008) & 0.624 (0.008) & 0.785 (0.019) & 0.722 (0.022) & 0.689 (0.025) \\
 500 & 0.719 (0.000) & 0.665 (0.000) & 0.631 (0.000) & 0.787 (0.016) & 0.727 (0.019) & 0.695 (0.021) \\
\hline
 Gaussian & 2.0 & 2.3 & 2.5 & 1.3 & 1.5 & 2.0 \\
\hline
 100 & 0.100 (0.008) & 0.079 (0.007) & 0.068 (0.007) & 0.427 (0.043) & 0.408 (0.040) & 0.400 (0.039) \\
 200 & 0.117 (0.007) & 0.095 (0.006) & 0.083 (0.005) & 0.438 (0.031) & 0.423 (0.026) & 0.415 (0.026) \\
 300 & 0.127 (0.005) & 0.103 (0.005) & 0.091 (0.004) & 0.440 (0.023) & 0.425 (0.023) & 0.422 (0.021) \\
 400 & 0.134 (0.004) & 0.110 (0.003) & 0.097 (0.003) & 0.441 (0.019) & 0.428 (0.017) & 0.423 (0.018) \\
 500 & 0.141 (0.000) & 0.116 (0.000) & 0.103 (0.000) & 0.442 (0.016) & 0.429 (0.015) & 0.424 (0.014) \\
\hline
    \end{tabular}
\end{table}


\section{Real Data Analysis}
Alzheimer's disease (AD) is the most common neurodegenerative disease and has substantial genetic components \citep{karch2014alzheimer, sims2020multiplex}. AD primarily targets the hippocampus region of the brain, which plays an important role in learning and memory. In the early stages of AD, the hippocampus suffers from volume loss \citep{schuff2009mri, mu2011adult}. Additionally, studies have shown that all hippocampal subregion volumes are highly heritable \citep{whelan2016heritability}. Accurately estimating genetic heritability can help us understand more about the genetic inheritance of AD and support the development of effective genetic treatments for the disease.

We applied our proposed genetic heritability estimator to whole genome sequencing data from the Alzheimer's Disease Neuroimaging Initiative (ADNI). A total of 808 samples from the screening and baseline of the ADNI1 and ADNI2 studies included whole genome sequencing data. After retaining SNPs that were called or imputed with $r^2>0.9$, we performed quality control by removing all SNPs with MAF <0.05 and any SNPs that did not pass the Hardy-Weinberg equilibrium test ($p$-value $< 1e-6$). Combining individuals having phenotype information, 576 individuals remain for the analysis. 

We began by regressing the natural logarithm of hippocampal volume on several common clinical covariates, including age, gender, education status, and the number of alleles in \textit{APOE}$\varepsilon$4. The residuals were then used as the response variable in the kernel ridge regression. The three kernel matrices used in the simulation studies—linear kernel, polynomial kernel of degree 2, and Gaussian kernel-were constructed based on the SNP matrix.

After applying our method to estimate the heritability based on the residuals using $n\lambda_n=0.8$, we obtained $\hat{H}^2=0.028$ for the linear kernel; $\hat{H}^2=0.115$ for the polynomial kernel and $\hat{H}^2=0.024$ for the Gaussian kernel. As in the simulation studies, the polynomial kernel gives the best performance among the three kernel matrices. We hypothesize that the reason of seeing these small numbers is that the age predictor has explained much of the variations in the response.

\section{Discussion}
In this paper, we proposed an estimator for the broad-sense genetic heritability of continuous phenotypes using kernel ridge regression. Theoretically, we provided an upper bound and a lower bound for the heritability estimator under certain regularity conditions. Consequently, we demonstrated that with an appropriate choice of the regularization parameter  $\lambda_n$ in the kernel ridge regression, the upper and lower bounds for the estimator encompass the underlying heritability. Simulation studies across various scenarios suggest that the proposed estimator can indeed capture the underlying heritability.

In the methodology section, we provided a general result on the bounds for our heritability estimator. It is worth noting that in some special cases, the bounds could be simpler. For instance, when $r:=\mrm{rank}(\mbf{K})=o(n)$, it then follows from the von Neumann trace inequality \citep{mirsky1975trace} that as $n\to\infty$,
\begin{align*}
    \frac{1}{n-1}I_2^g & =\frac{\sigma_\epsilon^2}{n-1}\mrm{tr}\left((\mbf{K}+n\lambda_n\mbf{I}_n)^{-1}\mbf{K}(\mbf{I}_n-\bar{\mbf{J}}_n)\mbf{K}(\mbf{K}+n\lambda_n\mbf{I}_n)^{-1}\right)+o_p(1)\\
    & =\frac{\sigma_\epsilon^2}{n-1}\mrm{tr}\left(\mbf{K}(\mbf{K}+n\lambda_n\mbf{I}_n)^{-2}\mbf{K}(\mbf{I}_n-\bar{\mbf{J}}_n)\right)+o_p(1)\\
    & \leq\frac{\sigma_\epsilon^2}{n-1}\sum_{i=1}^nl_i(\mbf{K}(\mbf{K}+n\lambda_n\mbf{I}_n)^{-2}\mbf{K})l_i(\mbf{I}_n-\bar{\mbf{J}}_n)\\
    & =\frac{\sigma_\epsilon^2}{n-1}\sum_{i=1}^r\left(\frac{l_i(\mbf{K})}{l_i(\mbf{K})+n\lambda_n}\right)^2+o_p(1)\\
    & \leq\frac{r\sigma_\epsilon^2}{n-1}+o_p(1)\\
    & =o_p(1).
\end{align*}
Such a case occurs when the kernel matrix is rank deficient, which is generally true in the low-dimensional case. Consequently, $\hat{\sigma}_g^2=\frac{1}{n-1}I_1^g+o_p(1)$. In other words, the integral term for the empirical spectral distribution of $\mbf{K}$ will vanish. 

Despite the popularity of the Gaussian kernel in practice, the results of the simulation studies and real data analysis suggest that the polynomial kernel generally provides the best performance. In reality, for a given dataset, there is little prior knowledge about which kernel matrix to use. As a future project, we will explore using machine learning techniques, such as neural networks, to learn from the data and self-determine the best kernel matrix. Additionally, estimators obtained from kernel ridge regression are usually biased, and this bias may deteriorate the performance of the proposed estimator. Whether debiasing techniques can improve the performance of heritability estimation will also be part of our future work.

\backmatter

\bmhead{Acknowledgements}

The authors would like to thank MathWorks at Texas State University for supporting this research.

\section*{Declarations}

\begin{itemize}
\item Funding: No funding was received for conducting this study.
\item Competing interests: The authors declare they have no financial interests. In terms of non-financial interests, Xiaoxi Shen has served as an editorial review board member for Frontiers in Systems Biology.
\item Ethics approval and consent to participate: Not applicable.
\item Consent for publication: Not applicable.
\item Data availability: The data used in the simulation studies can be found at \url{https://github.com/SxxMichael/KRR-Heritability}
\item Materials availability: Not applicable
\item Code availability: The code used to implement the methods can be found at \url{https://github.com/SxxMichael/KRR-Heritability}
\item Author contribution: Conceptualization: Olivia Bley, Elizabeth Lei, Andy Zhou; Methodology: Olivia Bley, Elizabeth Lei, Andy Zhou, Xiaoxi Shen; Formal analysis and investigation: Xiaoxi Shen; Writing - original draft preparation: Olivia Bley, Elizabeth Lei, Andy Zhou; Writing - review and editing: Olivia Bley, Elizabeth Lei, Andy Zhou, Xiaoxi Shen; Supervision: Xiaoxi Shen.
\end{itemize}

\noindent
If any of the sections are not relevant to your manuscript, please include the heading and write `Not applicable' for that section. 

\bigskip
\begin{flushleft}%
Editorial Policies for:

\bigskip\noindent
Springer journals and proceedings: \url{https://www.springer.com/gp/editorial-policies}

\bigskip\noindent
Nature Portfolio journals: \url{https://www.nature.com/nature-research/editorial-policies}

\bigskip\noindent
\textit{Scientific Reports}: \url{https://www.nature.com/srep/journal-policies/editorial-policies}

\bigskip\noindent
BMC journals: \url{https://www.biomedcentral.com/getpublished/editorial-policies}
\end{flushleft}

\newpage
\begin{appendices}

\section{Proofs}\label{secA1}

An appendix contains the proof of the main results in Section 2 of the main text.

\subsection{Proof of Proposition \ref{Prop: I3g and I3e}}
\begin{proof}
    Under (C1), it follows from Theorem 2.6.3 in \citet{vershynin2018high} that for any $t>0$,
    \begin{align*}
        \mbb{P}\left(\left.\abs{\frac{1}{n-1}I_3^g}>2t\right|\mcal{Z}\right) & \lesssim\exp\left\{-\frac{c(n-1)^2t^2}{C^2\norm{(\mbf{K}+n\lambda_n\mbf{I}_n)^{-1}\mbf{K}(\mbf{I}_n-\bar{\mbf{J}}_n)\mbf{K}(\mbf{K}+n\lambda_n\mbf{I}_n)^{-1}\mbf{g}}^2}\right\}.
    \end{align*}
    Since $\mbf{I}_n-\bar{\mbf{J}}_n$ is idempotent, it follows that $\norm{\mbf{I}_n-\bar{\mbf{J}}_n}_{op}=1$ and hence
    \begin{align*}
         & \norm{(\mbf{K}+n\lambda_n\mbf{I}_n)^{-1}\mbf{K}(\mbf{I}_n-\bar{\mbf{J}}_n)\mbf{K}(\mbf{K}+n\lambda_n\mbf{I}_n)^{-1}\mbf{g}}^2\\
          \leq & \norm{(\mbf{K}+n\lambda_n\mbf{I}_n)^{-1}\mbf{K}(\mbf{I}_n-\bar{\mbf{J}}_n)\mbf{K}(\mbf{K}+n\lambda_n\mbf{I}_n)^{-1}}_{op}^2\norm{\mbf{g}}^2\\
          \leq & \norm{\mbf{K}(\mbf{K}+n\lambda_n\mbf{I}_n)^{-2}\mbf{K}}_{op}\norm{\mbf{g}}^2\\
          \leq & \left(\max_{1\leq i\leq n}\frac{l_i(\mbf{K})}{l_i(\mbf{K})+n\lambda_n}\right)^4nM^2\\
          \leq & nM^2=\mcal{O}(n).
    \end{align*}
    Therefore,
    $$
    \mbb{P}\left(\left.\abs{\frac{1}{n-1}I_3^g}>2t\right|\mcal{Z}\right)\lesssim\exp\{-\mcal{O}(n)\}\to0,
    $$
    which implies that given $\mcal{Z}$, $\frac{1}{n-1}I_3^g=o_p(1)$. Similarly,
    $$
    \mbb{P}\left(\left.\abs{\frac{1}{n}I_3^\epsilon}>2t\right|\mcal{Z}\right)\lesssim\exp\left\{-\frac{cn^2t^2}{C^2\norm{n^2\lambda_n^2(\mbf{K}+n\lambda_n\mbf{I}_n)^{-2}\mbf{g}}^2}\right\}.
    $$
    Note that
    \begin{align*}
    \norm{n^2\lambda_n^2(\mbf{K}+n\lambda_n\mbf{I}_n)^{-2}\mbf{g}}^2 & \leq\norm{n^2\lambda_n^2(\mbf{K}+n\lambda_n\mbf{I}_n)^{-2}}_{op}^2\norm{\mbf{g}}^2\\
    & \leq\left(\max_{1\leq i\leq n}\left(\frac{n\lambda_n}{l_i(\mbf{K})+n\lambda_n}\right)^2\right)^2nM^2\\
    & \leq nM^2=\mcal{O}(n),
    \end{align*}
    Therefore, 
    $$
    \mbb{P}\left(\left.\abs{\frac{1}{n-1}I_3^\epsilon}>2t\right|\mcal{Z}\right)\lesssim\exp\{-\mcal{O}(n)\}\to0,
    $$
    which implies that given $\mcal{Z}$, $\frac{1}{n-1}I_3^\epsilon=o_p(1)$.
\end{proof}

\subsection{Proof of Proposition \ref{Prop: Asymptotics of I2}}
\begin{proof}
    Since $\epsilon_i$ are sub-gaussian random variables by (C1), it follows from the Hanson-Wright inequality \citep{rudelson2013hanson} that 
    $$
    \mbb{P}\left(\left.\abs{\frac{1}{n-1}(I_2^g-\mbb{E}[I_2^g])}>t\right|\mcal{Z}\right)\lesssim\exp\left\{-c\left[\frac{(n-1)^2t^2}{C^4\norm{\mbf{\Gamma}_1}_F^2}\wedge\frac{(n-1)t}{C^2\norm{\mbf{\Gamma}_1}_{op}}\right]\right\},
    $$
    where $\mbf{\Gamma}_1=(\mbf{K}+n\lambda_n\mbf{I}_n)^{-1}\mbf{K}(\mbf{I}_n-\bar{\mbf{J}}_n)\mbf{K}(\mbf{K}+n\lambda_n\mbf{I}_n)^{-1}$. Now note that
    \begin{align*}
        \norm{\mbf{\Gamma}_1}_{op} & \leq\norm{\mbf{K}(\mbf{K}+n\lambda_n\mbf{I}_n)^{-1}\mbf{K}}_{op}\norm{\mbf{I}_n-\bar{\mbf{J}}_n}_{op}\\
        & =\left(\frac{l_1(\mbf{K})}{l_1(\mbf{K})+n\lambda_n}\right)^2=\mcal{O}(1)\\
        \norm{\mbf{\Gamma}_1}_F^2 & =\mrm{tr}(\mbf{\Gamma}_1^2)\leq n\norm{\mbf{\Gamma}_1}_{op}^2=\mcal{O}(n),
    \end{align*}
    which implies that as $n\to\infty$,
    $$
    \mbb{P}\left(\left.\abs{\frac{1}{n-1}\left(I_2^g-\mbb{E}[I_2^g]\right)}>t\right|\mcal{Z}\right)\lesssim\exp\left\{-\mcal{O}(n)\right\}\to0.
    $$
    Therefore,
    \begin{align*}
        \frac{1}{n-1}I_2^g & =\frac{1}{n-1}\mbb{E}[I_2^g]+o_p(1)\\
        & =\frac{\sigma_\epsilon^2}{n-1}\mrm{tr}\left((\mbf{K}+n\lambda_n\mbf{I}_n)^{-1}\mbf{K}(\mbf{I}_n-\bar{\mbf{J}}_n)\mbf{K}(\mbf{K}+n\lambda_n\mbf{I}_n)^{-1}\right)+o_p(1).
    \end{align*}
    Similarly, we have
    $$
    \mbb{P}\left(\left.\abs{\frac{1}{n}(I_2^\epsilon-\mbb{E}[I_2^\epsilon])}>t\right|\mcal{Z}\right)\lesssim\exp\left\{-c\left[\frac{n^2t^2}{C^4\norm{\mbf{\Gamma}_2}_F^2}\wedge\frac{nt}{C^2\norm{\mbf{\Gamma}_2}_{op}}\right]\right\},
    $$
    where $\mbf{\Gamma}_2=n^2\lambda_n^2(\mbf{K}+n\lambda_n\mbf{I}_n)^{-2}$. Then note that
    \begin{align*}
        \norm{\mbf{\Gamma}_2}_{op} & =\left(\frac{n\lambda_n}{l_n(\mbf{K})+n\lambda_n}\right)^2=\mcal{O}(1).\\
        \norm{\mbf{\Gamma}_2}_F^2 & =\mrm{tr}(\mbf{\Gamma}_2^2)\leq n\norm{\mbf{\Gamma}_2}_{op}^2=\mcal{O}(n),
    \end{align*}
    which implies that as $n\to\infty$
    $$
    \mbb{P}\left(\left.\abs{\frac{1}{n}\left(I_2^\epsilon-\mbb{E}[I_2^\epsilon]\right)}>t\right|\mcal{Z}\right)\lesssim\exp\left\{-\mcal{O}(n)\right\}\to0.
    $$
    Therefore,
    \begin{align*}
        \frac{1}{n}I_2^\epsilon & =\frac{1}{n}\mbb{E}[I_2^\epsilon]+o_p(1)\\
        & =\frac{\sigma_\epsilon^2}{n}\mrm{tr}\left(n^2\lambda_n^2(\mbf{K}+n\lambda_n)^{-2}\right)+o_p(1).
    \end{align*}
\end{proof}

\subsection{Proof of Proposition \ref{Prop: Prelim Bounds for 1&g}}
\begin{proof}
    Note that
    \begin{align*}
        \abs{\mbf{1}_n^T\mbf{K}(\mbf{K}+n\lambda_n\mbf{I}_n)^{-1}\mbf{g}} & =\abs{\sum_{i=1}^n\frac{l_i(\mbf{K})}{l_i(\mbf{K})+n\lambda_n}(\mbf{v}_i^T\mbf{1}_n)(\mbf{v}_i^T\mbf{g})}\\
        & =\abs{\frac{l_1(\mbf{K})}{l_1(\mbf{K})+n\lambda_n}(\mbf{v}_1^T\mbf{1}_n)(\mbf{v}_1^T\mbf{g})+\sum_{i\geq2}\frac{l_i(\mbf{K})}{l_i(\mbf{K})+n\lambda_n}(\mbf{v}_i^T\mbf{1}_n)(\mbf{v}_i^T\mbf{g})}\\
        & \geq\frac{l_1(\mbf{K})}{l_1(\mbf{K})+n\lambda_n}\abs{\mbf{v_1^T\mbf{1}_n}}\abs{\mbf{v}_1^T\mbf{g}}-\sum_{i\geq2}\frac{l_i(\mbf{K})}{l_i(\mbf{K})+n\lambda_n}\abs{\mbf{v}_i^T\mbf{1}_n}\abs{\mbf{v}_i^T\mbf{g}}\\
        & \geq\left(\frac{c^2l_1(\mbf{K})}{l_1(\mbf{K})+n\lambda_n}-(1-c^2)\frac{l_2(\mbf{K})}{l_2(\mbf{K})+n\lambda_n}\right)\sqrt{n}\norm{\mbf{g}},
    \end{align*}
    where the last inequality follows since by assumption $\frac{l_1(\mbf{K})}{l_1(\mbf{K})+n\lambda_n}\abs{\mbf{v}_1^T\mbf{1}_n}\abs{\mbf{v}_1^T\mbf{g}}\geq\frac{l_1(\mbf{K})}{l_1(\mbf{K})+n\lambda_n} c^2\sqrt{n}\norm{\mbf{g}}$ and
    \begin{align*}
        \sum_{i\geq2}\frac{l_i(\mbf{K})}{l_i(\mbf{K})+n\lambda_n}\abs{\mbf{v}_i^T\mbf{1}_n}\abs{\mbf{v}_i^T\mbf{g}} & \leq\frac{l_2(\mbf{K})}{l_2(\mbf{K})+n\lambda_n}\sum_{i\geq2}\abs{\mbf{v}_i^T\mbf{1}_n}\abs{\mbf{v}_i^T\mbf{g}}\\
        & \leq\frac{l_2(\mbf{K})}{l_2(\mbf{K})+n\lambda_n}\sqrt{\sum_{i\geq 2}(\mbf{v}_i^T\mbf{1}_n)^2}\sqrt{\sum_{i\geq 2}(\mbf{v}_i^T\mbf{g})^2}\\
        & =\frac{l_2(\mbf{K})}{l_2(\mbf{K})+n\lambda_n}\sqrt{\sum_{i=1}^n(\mbf{v}_i^T\mbf{1}_n)^2-(\mbf{v}_1^T\mbf{1}_n)^2}\sqrt{\sum_{i=1}^n(\mbf{v}_i^T\mbf{g})^2-(\mbf{v}_1^T\mbf{g})^2}\\
        & \leq(1-c^2)\frac{l_2(\mbf{K})}{l_2(\mbf{K})+n\lambda_n}\sqrt{n}\norm{\mbf{g}}.
    \end{align*}
    Based on the choice of $n\lambda_n$ and (C4), it follows that $\frac{c^2l_1(\mbf{K})}{l_1(\mbf{K})+n\lambda_n}-(1-c^2)\frac{l_2(\mbf{K})}{l_2(\mbf{K})+n\lambda_n}\geq\alpha\left(\frac{l_2(\mbf{K})}{l_2(\mbf{K})+n\lambda_n}\right)$ and the desired result then follows.
\end{proof}

\subsection{Proof of Proposition \ref{Prop: Bounds for 1&g}}
\begin{proof}
    First, for the denominator $(\mbf{1}_n^T\mbf{g})^2$, we have
    $$
    c^2n\norm{\mbf{g}}^2\leq(\mbf{1}_n^T\mbf{g})^2=\mbf{g}^T\mbf{1}_n\mbf{1}_n^T\mbf{g}\leq\norm{\mbf{1}_n\mbf{1}_n^T}_{op}\norm{\mbf{g}}^2=n\norm{\mbf{g}}^2.
    $$
    On the other hand, we have
    \begin{align*}
        \frac{1}{n}(\mbf{1}_n^T\mbf{K}(\mbf{K}+n\lambda_n\mbf{I}_n)^{-1}\mbf{g})^2 & =\frac{1}{n}\mbf{g}^T(\mbf{K}+n\lambda_n\mbf{I}_n)^{-1}\mbf{K}\mbf{J}_n\mbf{K}(\mbf{K}+n\lambda_n\mbf{I}_n)^{-1}\mbf{g}\\
        & \leq\frac{1}{n}\norm{(\mbf{K}+n\lambda_n\mbf{I}_n)^{-1}\mbf{K}\mbf{J}_n\mbf{K}(\mbf{K}+n\lambda_n\mbf{I}_n)^{-1}}_{op}\norm{\mbf{g}}^2\\
        & \leq\frac{1}{n}\norm{\mbf{J}_n}\norm{\mbf{K}(\mbf{K}+n\lambda_n\mbf{I}_n)^{-2}\mbf{K}}_{op}\norm{\mbf{g}}^2\\
        & =\left(\frac{l_1(\mbf{K})}{l_1(\mbf{K})+n\lambda_n}\right)^2\norm{\mbf{g}}^2,
    \end{align*}
    and from Proposition \ref{Prop: Prelim Bounds for 1&g},
    \begin{align*}
        \frac{1}{n}(\mbf{1}_n^T\mbf{K}(\mbf{K}+n\lambda_n\mbf{I}_n)^{-1}\mbf{g})^2 & \geq\alpha^2\left(\frac{l_2(\mbf{K})}{l_2(\mbf{K})+n\lambda_n}\right)^2\norm{\mbf{g}}^2
    \end{align*}
    Hence, 
    $$
    \alpha^2\left(\frac{l_2(\mbf{K})}{l_2(\mbf{K})+n\lambda_n}\right)^2\leq \frac{\frac{1}{n}(\mbf{1}_n^T\mbf{K}(\mbf{K}+n\lambda_n\mbf{I}_n)^{-1}\mbf{g})^2}{\frac{1}{n}(\mbf{1}_n^T\mbf{g})^2}\leq c^{-2}\left(\frac{l_1(\mbf{K})}{l_1(\mbf{K})+n\lambda_n}\right)^2.
    $$
\end{proof}

\subsection{Proof of Proposition \ref{Prop: Bounds for I1g and I2g}}
\begin{proof}
    Note that
    \begin{align*}
        \frac{1}{n-1}I_1^g & =\frac{1}{n-1}\mbf{g}^T(\mbf{K}+n\lambda_n\mbf{I}_n)^{-1}\mbf{K}(\mbf{I}_n-\bar{\mbf{J}}_n)\mbf{K}(\mbf{K}+n\lambda_n\mbf{I}_n)^{-1}\mbf{g}\\
        & =\frac{1}{n-1}\left[\mbf{g}^T(\mbf{K}+n\lambda_n\mbf{I}_n)^{-1}\mbf{K}^2(\mbf{K}+n\lambda_n\mbf{I}_n)^{-1}\mbf{g}-\frac{1}{n}(\mbf{1}_n^T\mbf{K}(\mbf{K}+n\lambda_n\mbf{I}_n)^{-1}\mbf{g})^2\right]\\
        & =\frac{1}{n-1}\left[\sum_{i=1}^n\left(\frac{l_i(\mbf{K})}{l_i(\mbf{K})+n\lambda_n}\right)^2(\mbf{v}_i^T\mbf{g})^2-\frac{1}{n}(\mbf{1}_n^T\mbf{K}(\mbf{K}+n\lambda_n\mbf{I}_n)^{-1}\mbf{g})^2\right].
    \end{align*}
    Since by (C3),
    \begin{align*}
    c^2\left(\frac{l_1(\mbf{K})}{l_1(\mbf{K})+n\lambda_n}\right)^2\norm{\mbf{g}}^2 & \leq\left(\frac{l_1(\mbf{K})}{l_1(\mbf{K})+n\lambda_n}\right)^2(\mbf{v}_1^T\mbf{g})^2\\
    & \leq\sum_{i=1}^n\left(\frac{l_i(\mbf{K})}{l_i(\mbf{K})+n\lambda_n}\right)^2(\mbf{v}_i^T\mbf{g})^2\\
    & \leq\left(\frac{l_1(\mbf{K})}{l_1(\mbf{K})+n\lambda_n}\right)^2\norm{\mbf{g}}^2,
    \end{align*}
    it then follows from Proposition \ref{Prop: Bounds for 1&g} that
    \begin{align*}
        \frac{1}{n-1}I_1^g & \geq\frac{1}{n-1}\left[c^2\left(\frac{l_1(\mbf{K})}{l_1(\mbf{K})+n\lambda_n}\right)^2\norm{\mbf{g}}^2-c^{-2}\left(\frac{l_1(\mbf{K})}{l_1(\mbf{K})+n\lambda_n}\right)^2\frac{1}{n}(\mbf{1}_n^T\mbf{g})^2\right]\\
        & =\frac{nc^2}{n-1}\left(\frac{l_1(\mbf{K})}{l_1(\mbf{K})+n\lambda_n}\right)^2\left[\frac{1}{n}\norm{\mbf{g}}^2-c^{-4}\left(\frac{1}{n}\mbf{1}_n^T\mbf{g}\right)^2\right],
    \end{align*}
    and
    \begin{align*}
        \frac{1}{n-1}I_1^g & \leq\frac{1}{n-1}\left[\left(\frac{l_1(\mbf{K})}{l_1(\mbf{K})+n\lambda_n}\right)^2\norm{\mbf{g}}^2-\alpha^2\left(\frac{l_2(\mbf{K})}{l_2(\mbf{K})+n\lambda_n}\right)^2\frac{1}{n}(\mbf{1}_n^T\mbf{g})^2\right]\\
        & =\frac{n}{n-1}\left(\frac{l_1(\mbf{K})}{l_1(\mbf{K})+n\lambda_n}\right)^2\left[\frac{1}{n}\norm{\mbf{g}}^2-\alpha^{2}\left(\frac{l_2(\mbf{K})}{l_2(\mbf{K}+n\lambda_n)}\frac{l_1(\mbf{K})+n\lambda_n}{l_1(\mbf{K})}\right)^2\left(\frac{1}{n}\mbf{1}_n^T\mbf{g}\right)^2\right]
    \end{align*}
    Similarly, as a consequence of Proposition \ref{Prop: Asymptotics of I2}, the upper bound and lower bound of $\frac{1}{n-1}I_2^g$ can be obtained as follow:
    \begin{align*}
        \frac{1}{n-1}I_2^g & =\frac{\sigma_\epsilon^2}{n-1}\mrm{tr}\left((\mbf{K}+n\lambda_n)^{-1}\mbf{K}(\mbf{I}_n-\bar{\mbf{J}}_n)\mbf{K}(\mbf{K}+n\lambda_n\mbf{I}_n)^{-1}\right)+o_p(1)\\
        & =\frac{\sigma_\epsilon^2}{n-1}\left[\mrm{tr}\left(\mbf{K}(\mbf{K}+n\lambda_n\mbf{I}_n)^{-2}\mbf{K}\right)-\frac{1}{n}\mbf{1}_n^T\mbf{K}(\mbf{K}+n\lambda_n\mbf{I}_n)^{-2}\mbf{K}\mbf{1}_n\right]+o_p(1)\\
        & =\frac{\sigma_\epsilon^2}{n-1}\left[\sum_{i=1}^n\left(\frac{l_i(\mbf{K})}{l_i(\mbf{K})+n\lambda_n}\right)^2-\frac{1}{n}\sum_{i=1}^n\left(\frac{l_i(\mbf{K})}{l_i(\mbf{K})+n\lambda_n}\right)^2(\mbf{v}_i^T\mbf{1}_n)^2\right]+o_p(1)\\
        & \leq\frac{(1-c^2)\sigma_\epsilon^2}{n-1}\sum_{i=1}^n\left(\frac{l_i(\mbf{K})}{l_i(\mbf{K})+n\lambda_n}\right)^2+o_p(1),
    \end{align*}
    where the last inequality follows from (C3). On the other hand, since $\mbf{1}_n^T\mbf{K}(\mbf{K}+n\lambda_n\mbf{I}_n)^{-2}\mbf{K}\mbf{1}_n\leq n\left(\frac{l_1(\mbf{K})}{l_1(\mbf{K})+n\lambda_n}\right)^2$, it then follows that
    $$
    \frac{1}{n-1}I_2^g\geq\frac{\sigma_\epsilon^2}{n-1}\sum_{i=2}^n\left(\frac{l_i(\mbf{K})}{l_i(\mbf{K})+n\lambda_n}\right)^2+o_p(1).
    $$
\end{proof}

\subsection{Proof of Proposition \ref{Prop: Bounds for I1e and I2e}}
\begin{proof}
    Note that
    $$
    \frac{1}{n}I_1^\epsilon=\frac{1}{n}\sum_{i=1}^n\left(\frac{n\lambda_n}{l_i(\mbf{K})+n\lambda_n}\right)^2(\mbf{v}_i^T\mbf{g})^2,
    $$
    and since
    \begin{align*}
        \sum_{i=1}^n\left(\frac{n\lambda_n}{l_i(\mbf{K})+n\lambda_n}\right)^2(\mbf{v}_i^T\mbf{g})^2 & \geq\left(\frac{n\lambda_n}{l_1(\mbf{K})+n\lambda_n}\right)^2\norm{\mbf{g}}^2\\
        \sum_{i=1}^n\left(\frac{n\lambda_n}{l_i(\mbf{K})+n\lambda_n}\right)^2(\mbf{v}_i^T\mbf{g})^2 & \leq\left(\frac{n\lambda_n}{l_n(\mbf{K})+n\lambda_n}\right)^2\norm{\mbf{g}}^2\leq\norm{\mbf{g}}^2
    \end{align*}
    it then follows that
    $$
    \frac{1}{n}\left(\frac{n\lambda_n}{l_1(\mbf{K})+n\lambda_n}\right)^2\norm{\mbf{g}}^2\leq\frac{1}{n}I_1^\epsilon\leq\frac{1}{n}\left(\frac{n\lambda_n}{l_n(\mbf{K})+n\lambda_n}\right)^2\norm{\mbf{g}}^2\leq\frac{1}{n}\norm{\mbf{g}}^2.
    $$
    Similarly, as a consequence of Proposition \ref{Prop: Asymptotics of I2}, we have
    \begin{align*}
        \frac{1}{n}I_2^\epsilon & =\frac{\sigma_\epsilon^2}{n}\mrm{tr}\left(n^2\lambda_n^2(\mbf{K}+n\lambda_n\mbf{I}_n)^{-2}\right)+o_p(1)\\
        & =\frac{\sigma_\epsilon^2}{n}\sum_{i=1}^n\left(\frac{n\lambda_n}{l_i(\mbf{K})+n\lambda_n}\right)^2+o_p(1)\\
        & \leq\left(\frac{n\lambda_n}{l_n(\mbf{K}+n\lambda_n)}\right)^2\sigma_\epsilon^2+o_p(1)\\
        & \leq\sigma_\epsilon^2+o_p(1),
    \end{align*}
    and
    $$
    \frac{1}{n}I_2^\epsilon\geq\left(\frac{n\lambda_n}{l_1(\mbf{K})+n\lambda_n}\right)^2\sigma_\epsilon^2+o_p(1).
    $$
\end{proof}

\subsection{Proof of Proposition \ref{Prop: Bounds for sigmae^2/sigmag^2}}
\begin{proof}
    The upper bound and the lower bound for $\hat{\sigma}_\epsilon^2/\hat{\sigma}_g^2$ follow directly from Theorem \ref{Thm: Bounds for sigmag and sigmae}. In addition, denote LB as the lower bound of $\hat{\sigma}_\epsilon^2/\hat{\sigma}_g^2$ and UB as the upper bound of $\hat{\sigma}_\epsilon^2/\hat{\sigma}_g^2$ for simplicity. Note that $c^2\tau_1^2-c^{-4}\leq0$,
    $$
    UB\geq\frac{\sigma_\epsilon^2}{c^2\tau_1^2\sigma_g^2+\sigma_\epsilon^2\int\left(\frac{x}{x+n\lambda_n}\right)^2\mrm{d}F_{-1}(x)},
    $$
    and based on the choice of $\lambda_n$, it follows that 
    $$
    UB\geq\frac{\sigma_\epsilon^2}{c^2\tau_1^2\sigma_g^2+\sigma_\epsilon^2\int\left(\frac{x}{x+n\lambda_n}\right)^2\mrm{d}F_{-1}(x)}\geq\frac{\sigma_\epsilon^2}{\sigma_g^2}.
    $$
    Similarly, based on the choice of $\lambda_n$, we have
    $$
    \frac{\sigma_\epsilon^2}{\sigma_g^2} \geq(1-\tau_1^2)^{-1}LB\geq LB.
    $$
\end{proof}

\newpage

\section{Additional Simulation Results}

\begin{table}[htbp]
    \centering
    \caption{Mean and standard deviation of the heritability estimators obtained after running 500 repetitions when the underlying function is quadratic and the low dimensional case was applied. The true heritability for the HWE scenario is 0.573 and the true heritability for the 1000 Genome Project case is 0.811.}
    \begin{tabular}{c|ccc|ccc}
\hline 
 & \multicolumn{3}{c|}{Hardy Weinberg Equilibrium}
 & \multicolumn{3}{c}{1000 Genome Project} \\
\hline
 & \multicolumn{3}{c|}{$n\lambda_n$}
 & \multicolumn{3}{c}{$n\lambda_n$} \\
\hline
 Linear & 2.3 & 2.5 & 3.0 & 0.5 & 0.8 & 1 \\
\hline
 600 & 0.125 (0.017) & 0.110 (0.015) & 0.082 (0.011) & 0.836 (0.071) & 0.756 (0.080) & 0.706 (0.087) \\
 700	& 0.122 (0.015) & 0.106 (0.013) & 0.081 (0.010) & 0.823 (0.050) & 0.749 (0.057) & 0.698 (0.065) \\
 800	&	0.117 (0.012) &	0.104 (0.011) &	0.080 (0.008) & 0.807 (0.039) & 0.733 (0.047) & 0.685 (0.053) \\
 900	& 0.112 (0.010) & 0.102 (0.009) & 0.079 (0.008) & 0.792 (0.034) & 0.712 (0.041) &	0.670 (0.042) \\
 1000 &	0.109 (0.009) &	0.098 (0.008) &	0.077 (0.007) & 0.776 (0.028) & 0.696 (0.034) &	0.653 (0.037) \\
\hline
 Polynomial & 2.3 & 2.5 & 3.0 & 0.5 & 0.8 & 1 \\
\hline
 600 & 0.608 (0.028) & 0.571 (0.029) & 0.486 (0.030) & 0.948 (0.030) & 0.926 (0.039) & 0.912 (0.044) \\
 700	& 0.583 (0.024) & 0.543 (0.024) & 0.463 (0.026) & 0.942 (0.023) & 0.924 (0.026) & 0.910 (0.030) \\
 800	&	0.558 (0.023) &	0.523 (0.022) &	0.444 (0.022) &	0.937 (0.017) & 0.918 (0.022) &	0.904 (0.026) \\
 900	& 0.536 (0.019) & 0.504 (0.019) & 0.425 (0.021) &	0.933 (0.015) &	0.910 (0.019) &	0.901 (0.021) \\
 1000 &	0.517 (0.017) &	0.482 (0.017) &	0.410 (0.017) & 0.929 (0.012) & 0.907 (0.016) &	0.896 (0.018) \\
\hline
 Gaussian & 2.3 & 2.5 & 3.0 & 0.5 & 0.8 & 1 \\
\hline
 600 & 0.072 (0.005) & 0.062 (0.005) &	0.045 (0.003) & 0.649 (0.058) & 0.450 (0.047) & 0.352 (0.042) \\
 700	& 0.072 (0.005) & 0.062 (0.004) &	0.045 (0.003) & 0.650 (0.039) & 0.458 (0.033) & 0.361 (0.031) \\
 800	&	0.071 (0.005) &	0.062 (0.004) &	0.045 (0.003) &	0.652 (0.031) & 0.463 (0.030) &	0.368 (0.028)\\
 900	& 0.071 (0.004) & 0.062 (0.004) & 0.045 (0.003) &	0.653 (0.027) &	0.464 (0.028) &	0.372 (0.025) \\
 1000 &	0.071 (0.004) &	0.061 (0.003) &	0.045 (0.003) & 0.652 (0.023) & 0.466 (0.024) &	0.375 (0.023)\\
\hline
    \end{tabular}
\end{table}

\begin{table}[htbp]
\centering
\caption{Mean and standard deviation of the heritability estimators obtained after running 500 repetitions when the underlying function is quadratic and the high dimensional case was applied. The true heritability for the HWE scenario is 0.602 and the true heritability for the 1000 Genome Project case is 0.553.}
    \begin{tabular}{c|ccc|ccc}
\hline
 & \multicolumn{3}{c|}{Hardy Weinberg Equilibrium}
 & \multicolumn{3}{c}{1000 Genome Project} \\
\hline
 & \multicolumn{3}{c|}{$n\lambda_n$}
 & \multicolumn{3}{c}{$n\lambda_n$} \\
\hline
 Linear & 2.0 & 2.3 & 2.5 & 1.5 & 2.0 & 2.3 \\
\hline
 100 & 0.152 (0.014) & 0.122 (0.015) & 0.106 (0.010) & 0.381 (0.077) & 0.292 (0.072) & 0.248 (0.069) \\
 200 & 0.145 (0.011) & 0.117 (0.009) & 0.106 (0.010) & 0.376 (0.059) & 0.294 (0.053) & 0.250 (0.055) \\
 300 & 0.137 (0.009) & 0.112 (0.007) & 0.099 (0.007) & 0.372 (0.046) & 0.291 (0.044) & 0.257 (0.042) \\
 400 & 0.129 (0.006) & 0.106 (0.005) & 0.094 (0.005) & 0.367 (0.040) & 0.288 (0.036) & 0.253 (0.035) \\
 500 & 0.119 (9.45E-17) & 0.099 (8.07E-17) & 0.088 (7.76E-17) & 0.356 (0.031) & 0.283 (0.031) & 0.251 (0.029) \\
\hline
 Polynomial & 2.0 & 2.3 & 2.5 & 1.5 & 2.0 & 2.3 \\
\hline
 100 & 0.679 (0.014) & 0.618 (0.015) & 0.580 (0.015) & 0.770 (0.063) & 0.705 (0.081) & 0.661 (0.089) \\
 200 & 0.666 (0.013) & 0.605 (0.013) & 0.567 (0.013) & 0.764 (0.049) & 0.701 (0.054) & 0.658 (0.066) \\
 300 & 0.656 (0.010) & 0.594 (0.011) & 0.556 (0.012) & 0.761 (0.038) & 0.695 (0.045) & 0.664 (0.049) \\
 400 & 0.64 (0.008) & 0.583 (0.009) & 0.546 (0.008) & 0.755 (0.033) & 0.692 (0.036) & 0.656 (0.038) \\
 500 & 0.635 (3.10E-16) & 0.574 (3.39E-16) & 0.537 (3.47E-16) & 0.749 (0.026) & 0.687 (0.031) & 0.653 (0.032) \\
\hline
 Gaussian & 2.0 & 2.3 & 2.5 & 1.5 & 2.0 & 2.3 \\
\hline
 100 & 0.091 (0.004) & 0.071 (0.003) & 0.061 (0.003) & 0.154 (0.023) & 0.096 (0.016) & 0.074 (0.012) \\
 200 & 0.091 (0.004) & 0.071 (0.003) & 0.061 (0.003) & 0.162 (0.020) & 0.102 (0.014) & 0.079 (0.011) \\
 300 & 0.091 (0.003) & 0.071 (0.003) & 0.061 (0.002) & 0.169 (0.018) & 0.108 (0.013) & 0.086 (0.010) \\
 400 & 0.091 (0.003) & 0.071 (0.002) & 0.061 (0.002) & 0.177 (0.017) & 0.113 (0.012) & 0.090 (0.010) \\
 500 & 0.091 (1.38E-16) & 0.071 (1.15E-16) & 0.062 (1.13E-16) & 0.179 (0.014) & 0.118 (0.011) & 0.095 (0.010) \\
\hline
\end{tabular}
\end{table}

\begin{table}[htbp]
\centering
\caption{Mean and standard deviation of the heritability estimators obtained after running 500 repetitions when the underlying function is trigonometric and the low dimensional case was applied. The true heritability for the HWE scenario is 0.706 and the true heritability for the 1000 Genome Project case is 0.686.}
    \begin{tabular}{c|ccc|ccc}
\hline
 & \multicolumn{3}{c|}{Hardy Weinberg Equilibrium}
 & \multicolumn{3}{c}{ 1000 Genome Project} \\
\hline
 & \multicolumn{3}{c|}{$n\lambda_n$}
 & \multicolumn{3}{c}{$n\lambda_n$} \\
\hline
 Linear & 1.0 & 1.3 & 1.5 & 1.3 & 1.5 & 2.0 \\
\hline
 100 & 0.648 (0.050) & 0.583 (0.055) &	0.499 (0.064) & 0.395 (0.021) & 0.326 (0.020) & 0.218 (0.015)\\
 200	& 0.682 (0.035) & 0.640 (0.037) &	0.530 (0.047) & 0.411 (0.018) & 0.343 (0.017) & 0.233 (0.013) \\
 300	&	0.703 (0.026) &	0.678 (0.026)&	0.559 (0.033)&	0.428 (0.015) & 0.355 (0.015) &	0.247 (0.012)\\
 400	& 0.719 (0.020)& 0.704 (0.020)& 0.581 (0.028)&	0.444 (0.011)&	0.369 (0.010)&	0.263 (0.008)\\
 500 &	0.731 (0.017) &	0.721 (0.016) &	0.601 (0.023) &0.458 (0.000) &0.382 (0.000) &	0.277 (0.000)\\
\hline
 Polynomial & 1.0 & 1.3 & 1.5 & 1.3 & 1.5 & 2.0 \\
\hline
 100 & 0.903 (0.021) & 0.869 (0.023) &	0.846 (0.037) & 0.848 (0.009) & 0.805 (0.011) & 0.701 (0.014)\\
 200	& 0.914 (0.014) & 0.890 (0.016) &	0.857 (0.024) & 0.850 (0.007) & 0.808 (0.009) & 0.706 (0.016) \\
 300	&	0.920 (0.010) &	0.902 (0.010)&	0.870 (0.015)&	0.874 (0.006)& 0.810 (0.007) &	0.612 (0.010)\\
 400	& 0.924 (0.007) & 0.911 (0.007)& 0.878 (0.012)&	0.857 (0.004)&	0.813 (0.005)&	0.719 (0.006)\\
 500 &	0.928 (0.006) &	0.917 (0.006) &	0.886 (0.009) &0.860 (0.000) &0.816 (0.000) &	0.725 (0.000)\\
\hline
 Gaussian & 1.0 & 1.3 & 1.5 & 1.3 & 1.5 & 2.0 \\
\hline
 100 & 0.145 (0.021) & 0.107 (0.17) &	0.077 (0.013) & 0.203 (0.001) & 0.161 (0.001)& 0.099 (0.000)\\
 200	& 0.188 (0.020) & 0.144 (0.018) &	0.098 (0.014) & 0.212 (0.001) & 0.170 (0.000) & 0.107 (0.000) \\
 300	&	0.226 (0.019)&	0.174 (0.018)&	0.117 (0.009)&	0.221 (0.001)& 0.179 (0.000) &	0.115 (0.000)\\
 400	& 0.259 (0.019)& 0.203 (0.018)& 0.134 (0.009)&	0.232 (0.001)&	0.188 (0.000)&	0.123 (0.000)\\
 500 &	0.287 (0.017) &	0.226 (0.017) &	0.151 (0.009) &0.241 (0.000) &0.197 (0.000) &	0.130 (0.000)\\
\hline
\end{tabular}
\end{table}

\begin{table}[htbp]
\centering
\caption{Mean and standard deviation of the heritability estimators obtained after running 500 repetitions when the underlying function is trigonometric and the high dimensional case was applied. The true heritability for the HWE scenario is 0.724 and the true heritability for the 1000 Genome Project case is 0.767.}
    \begin{tabular}{c|ccc|ccc}
\hline
 & \multicolumn{3}{c|}{Hardy Weinberg Equilibrium}
 & \multicolumn{3}{c}{1000 Genome Project} \\
\hline
 & \multicolumn{3}{c|}{$n\lambda_n$}
 & \multicolumn{3}{c}{$n\lambda_n$} \\
\hline
 Linear & 1.3 & 1.5 & 2.0 & 1.3 & 1.5 & 2.0 \\
\hline
 100 & 0.544 (0.050) & 0.475 (0.067) &	0.365 (0.046) & 0.384 (0.020) & 0.323 (0.018) & 0.216 (0.014)\\
 200	& 0.581 (0.038) & 0.508 (0.045) &	0.404 (0.034) & 0.396 (0.018) & 0.338 (0.016) & 0.230 (0.014) \\
 300	& 0.606 (0.026)&	0.532 (0.034)&	0.437 (0.024)&	0.409 (0.014)& 0.352 (0.013) &	0.245 (0.013)\\
 400	& 0.624 (0.021)& 0.556 (0.028)& 0.459 (0.026)&	0.419 (0.011)&	0.366 (0.010)&	0.257 (0.009)\\
 500 & 0.642 (0.018) &0.0574 (0.027) &	0.479 (0.019) &0.430 (0.021) &0.380 (0.000) &	0.270 (0.000)\\
\hline
 Polynomial & 1.3 & 1.5 & 2.0 & 1.3 & 1.5 & 2.0 \\
\hline
 100 & 0.857 (0.028) & 0.831 (0.042) &	0.777 (0.040) & 0.843 (0.008) & 0.803 (0.009) & 0.699 (0.013)\\
 200	& 0.869 (0.019) & 0.848 (0.025) &	0.800 (0.025) & 0.843 (0.007) & 0.805 (0.008) & 0.703 (0.014) \\
 300	&	0.879 (0.012)&	0.858 (0.018)&	0.817 (0.016)&	0.845 (0.006)& 0.808 (0.007) &	0.708 (0.009)\\
 400	& 0.885 (0.010)& 0.869 (0.014)& 0.827 (0.013)&	0.846 (0.004)&	0.811 (0.005)&	0.712 (0.007)\\
 500 &	0.891 (0.007) &	0.876 (0.010) &	0.836 (0.011) &0.847 (0.000) &0.815 (0.000) &	0.717 (0.000)\\
\hline
 Gaussian & 1.3 & 1.5 & 2.0 & 1.3 & 1.5 & 2.0 \\
\hline
 100 & 0.078 (0.014) & 0.054 (0.012) &	0.032 (0.006) & 0.204 (0.001) & 0.161 (0.000) & 0.098 (0.000)\\
 200	& 0.103 (0.015) & 0.064 (0.010) &	0.0425 (0.006) & 0.215 (0.001) & 0.170 (0.000) & 0.104 (0.000) \\
 300	&	0.125 (0.015)&	0.073 (0.009)&	0.053 (0.006)&	0.225 (0.001)& 0.179 (0.000) &	0.110 (0.000)\\
 400	& 0.144 (0.014)& 0.082 (0.008)& 0.064 (0.007)&	0.237 (0.000)&	0.188 (0.000)&	0.118 (0.000)\\
 500 &	0.162 (0.014) &	0.091 (0.008) &	0.073 (0.007) &0.247 (0.000) &0.197 (0.000) &	0.124 (0.000)\\
\hline
\end{tabular}
\end{table}




\end{appendices}

\newpage
\bibliography{krrheritability}

\end{document}